\documentclass[superscriptaddress,amsmath,amssymb,aps,prm,twocolumn,nofootinbib,longbibliography]{revtex4-2}
\usepackage{graphicx}         

\usepackage{textcmds}
\usepackage{dcolumn}          
\usepackage{bm}               
\usepackage{upgreek}
\usepackage{booktabs} 				
\usepackage{tabularx}
\usepackage{array}
\usepackage[colorlinks=true,urlcolor=blue,breaklinks=true]{hyperref}
\usepackage{xcolor}
\usepackage[a4paper,left=2cm,right=2cm]{geometry}
\usepackage{orcidlink}
\usepackage{gensymb}
\usepackage{bibunits}
\usepackage{titlesec}

\begin{document}

\title{Digitally-alloyed magnetic topological insulator nanostructures grown via stencil mask molecular beam epitaxy}

\author{Jan Karthein\,\orcidlink{0009-0009-6887-8016}}
\thanks{These authors contributed equally to this work}
\affiliation{Peter Gr\"unberg Institut (PGI-9), Forschungszentrum J\"ulich, 52425 J\"ulich, Germany}
\affiliation{JARA-Fundamentals of Future Information Technology, J\"ulich-Aachen Research Alliance, Forschungszentrum J\"ulich and RWTH Aachen University, 52425 J\"ulich, Germany}

\author{Max Va{\ss}en-Carl\,\orcidlink{0000-0003-0895-7176}}
\thanks{These authors contributed equally to this work}
\affiliation{Peter Gr\"unberg Institut (PGI-9), Forschungszentrum J\"ulich, 52425 J\"ulich, Germany}
\affiliation{JARA-Fundamentals of Future Information Technology, J\"ulich-Aachen Research Alliance, Forschungszentrum J\"ulich and RWTH Aachen University, 52425 J\"ulich, Germany}

\author{Lars Maier\,\orcidlink{0009-0005-5887-1671}}
\affiliation{Peter Gr\"unberg Institut (PGI-9), Forschungszentrum J\"ulich, 52425 J\"ulich, Germany}
\affiliation{JARA-Fundamentals of Future Information Technology, J\"ulich-Aachen Research Alliance, Forschungszentrum J\"ulich and RWTH Aachen University, 52425 J\"ulich, Germany}

\author{Abdur Rehman Jalil\,\orcidlink{0000-0003-1869-2466}}
\affiliation{Peter Gr\"unberg Institut (PGI-9), Forschungszentrum J\"ulich, 52425 J\"ulich, Germany}
\affiliation{Institute of Experimental Physics III, University of W\"urzburg, 97070 W\"urzburg, Germany}

\author{Gion Toehgiono}
\affiliation{Peter Gr\"unberg Institut (PGI-9), Forschungszentrum J\"ulich, 52425 J\"ulich, Germany}
\affiliation{JARA-Fundamentals of Future Information Technology, J\"ulich-Aachen Research Alliance, Forschungszentrum J\"ulich and RWTH Aachen University, 52425 J\"ulich, Germany}

\author{Josua Thieme\,\orcidlink{0009-0008-5538-6891}}
\affiliation{Peter Gr\"unberg Institut (PGI-9), Forschungszentrum J\"ulich, 52425 J\"ulich, Germany}
\affiliation{JARA-Fundamentals of Future Information Technology, J\"ulich-Aachen Research Alliance, Forschungszentrum J\"ulich and RWTH Aachen University, 52425 J\"ulich, Germany}

\author{Julian Feldewerth\,\orcidlink{0009-0000-7226-664X}}
\affiliation{Peter Gr\"unberg Institut (PGI-9), Forschungszentrum J\"ulich, 52425 J\"ulich, Germany}
\affiliation{JARA-Fundamentals of Future Information Technology, J\"ulich-Aachen Research Alliance, Forschungszentrum J\"ulich and RWTH Aachen University, 52425 J\"ulich, Germany}

\author{Benjamin Bennemann\,\orcidlink{0000-0003-4071-019X}}
\affiliation{Peter Gr\"unberg Institut (PGI-10), Forschungszentrum J\"ulich, 52425 J\"ulich, Germany}
\affiliation{JARA-Fundamentals of Future Information Technology, J\"ulich-Aachen Research Alliance, Forschungszentrum J\"ulich and RWTH Aachen University, 52425 J\"ulich, Germany}

\author{Joscha Domnick}
\affiliation{Peter Gr\"unberg Institut (PGI-9), Forschungszentrum J\"ulich, 52425 J\"ulich, Germany}
\affiliation{JARA-Fundamentals of Future Information Technology, J\"ulich-Aachen Research Alliance, Forschungszentrum J\"ulich and RWTH Aachen University, 52425 J\"ulich, Germany}

\author{Taizo Kawano}
\affiliation{Department of Material Science, Tohoku University, Aobayama 2, Aoba-ku, Sendai 980-8579, Japan}

\author{Sosuke Otsubo}
\affiliation{Department of Material Science, Tohoku University, Aobayama 2, Aoba-ku, Sendai 980-8579, Japan}

\author{Ren\'e Heller\,\orcidlink{0000-0002-7679-3401}}
\affiliation{Helmholtz-Zentrum Dresden-Rossendorf, Institute of Ion Beam Physics and Materials Research, 01328 Dresden, Germany}

\author{Jun Ishihara\,\orcidlink{0000-0002-9574-1332}}
\affiliation{Department of Material Science, Tohoku University, Aobayama 2, Aoba-ku, Sendai 980-8579, Japan}

\author{Detlev Gr\"utzmacher\,\orcidlink{0000-0001-6290-9672}}
\affiliation{Peter Gr\"unberg Institut (PGI-9), Forschungszentrum J\"ulich, 52425 J\"ulich, Germany}
\affiliation{JARA-Fundamentals of Future Information Technology, J\"ulich-Aachen Research Alliance, Forschungszentrum J\"ulich and RWTH Aachen University, 52425 J\"ulich, Germany}

\author{Makoto Kohda\,\orcidlink{0000-0002-1995-1282}}
\affiliation{Department of Material Science, Tohoku University, Aobayama 2, Aoba-ku, Sendai 980-8579, Japan}

\author{Peter Sch\"uffelgen\,\orcidlink{0000-0001-7977-7848}}
\affiliation{Peter Gr\"unberg Institut (PGI-9), Forschungszentrum J\"ulich, 52425 J\"ulich, Germany}

\author{Thomas Sch\"apers\,\orcidlink{0000-0001-7861-5003}}
\email{th.schaepers@fz-juelich.de}
\affiliation{Peter Gr\"unberg Institut (PGI-9), Forschungszentrum J\"ulich, 52425 J\"ulich, Germany}
\affiliation{JARA-Fundamentals of Future Information Technology, J\"ulich-Aachen Research Alliance, Forschungszentrum J\"ulich and RWTH Aachen University, 52425 J\"ulich, Germany}
\affiliation{Department of Material Science, Tohoku University, Aobayama 2, Aoba-ku, Sendai 980-8579, Japan}

\hyphenation{}
\date{\today}

\begin{abstract}
Magnetic topological insulator nanostructures are a promising platform for the realization of fault-tolerant quantum computation. For this, such nanostructures need to be embedded into the necessary superconducting architectures. In this study, we demonstrate a method for the growth of magnetic topological insulator nanostructures via molecular beam epitaxy, utilising digital alloying and defining their shapes in-situ by stencil mask technology. Cr-doped (Bi$_y$Sb$_{1-y}$)$_2$Te$_3$ micro- and nanometer-sized Hall bars are fabricated and the magneto-transport properties for various Cr doping amounts and different Hall bar dimensions are investigated at low temperatures. We find indications that the Fermi energy is located inside the bulk band gap, while disorder and charge puddles affect the transport behaviour. Furthermore, we investigate the magnetic properties and discuss the transition from ferromagnetic to paramagnetic behaviour.
\end{abstract}

\maketitle

\section{Introduction}
\begin{bibunit}[]

\begin{figure*}[!t]
\centering
\includegraphics[width=\textwidth]{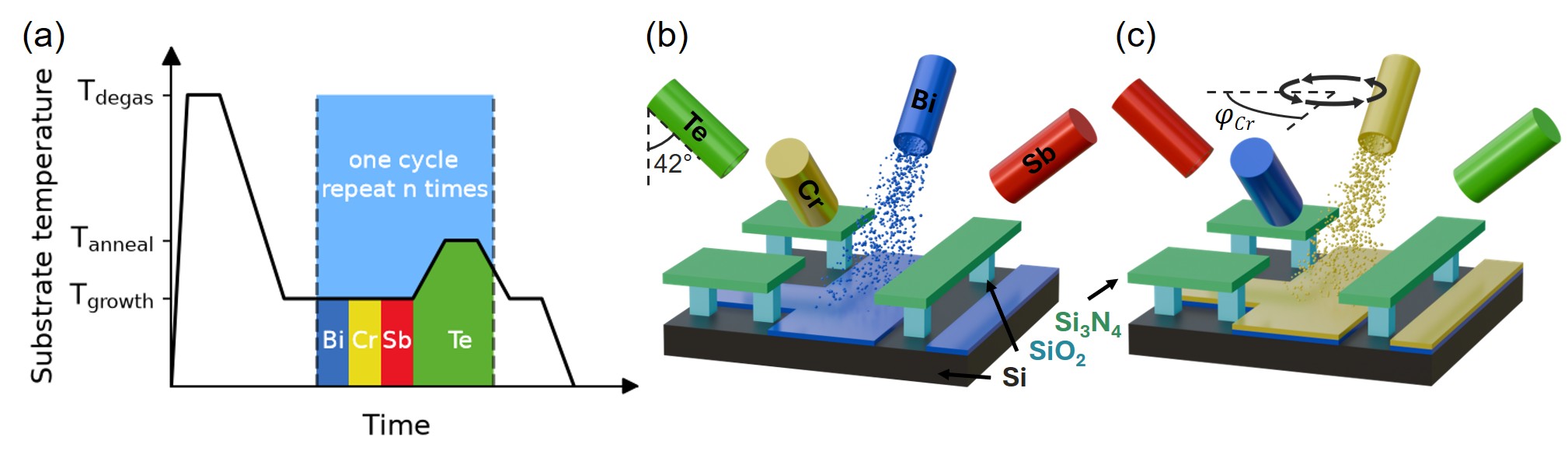}
\caption{Illustration of the digital alloying process. (a) Substrate temperature versus time diagram showing the order in which the elements are deposited.  (b) Schematic illustration of the Bi deposition step. The stencil mask is designed to interrupt the film in certain areas by the shadow of the mask and by that create arbitrary shapes. (c) Subsequent Cr deposition step. The substrate is rotated to a specific azimuthal angle $\varphi_\text{Cr}$ so that Cr is deposited from the same direction as Bi previously. Thus, the angle of each cell with respect to the substrate is identical. For clarity, in the figure the cells are rotated for illustrative purpose, whereas in our actual setup the substrate is rotated while the cells are stationary.}
\label{Figure1}
\end{figure*}

Magnetic topological insulators (MTIs) are a promising material platform for  realizing fault-tolerant quantum computation utilizing  Majorana zero modes~\cite{Chen2018,Qi2010,Tokura2019}. To achieve this, the chiral edge channels of an MTI nanoribbon need to be proximitized with an s-wave superconductor~\cite{Burke2024,Zsurka2025}. There are already some reported cases where superconducting correlations are present in antiferromagnetic~\cite{Jansen2024,Zhang2025} and ferromagnetic MTIs~\cite{Uday2024,Mandal2022}, but theoretical studies motivate to explore lower magnetic doping for superconducting applications~\cite{Legendre2024,Zsurka2025,Moors2025,Burke2024}. A prominent material is Cr-doped (Bi$_y$Sb$_{1-y}$)$_2$Te$_3$. Previously, it has been studied how the ferromagnetic and magneto-transport properties of this type of MTI change as a result of varying the Cr content~\cite{Kou2013,Chang2013,Ou2016}. Therefore, in a previous work we investigated low Cr-doped magnetic topological insulator thin films grown via molecular beam epitaxy (MBE) and showed that even for weak internal magnetization, a pronounced anomalous Hall effect is observed~\cite{Karthein2025}. This means that we are able to tune the MTI into a regime that is very promising for superconducting applications~\cite{Burke2024,Zsurka2025}. A challenge of embedding MTI nanoribbons in superconducting architectures is the need for very clean interfaces. For regular topological insulator nanostructures grown via MBE, a complete in-situ process has proven to be superior to ex-situ processes~\cite{Schueffelgen2019}. It also has been reported that electron beam lithography as a tool to create nanostructures can cause damage to the MTI material~\cite{Qiu2022,Andersen2023,Rosen2024}.

In order to tackle both problems at once, we developed a fully in-situ process scheme to fabricate MTI nanostructures by using MBE and stencil-masks. In order to demonstrate the feasibility of this processing scheme, Hall bars are fabricated in varying dimensions ranging from micro- to nanometer-sized devices. Furthermore, we investigate how Cr content affects the magnetic and electric properties at low temperatures, and discuss how narrowing down the width of the Hall bars changes these properties. The latter is particularly interesting because narrow MTI nanoribbons play a central role in proposed Majorana architectures.    

\section{Methods}

\subsection{Digital alloying}

\begin{figure*}
\centering
\includegraphics[width=0.8\textwidth]{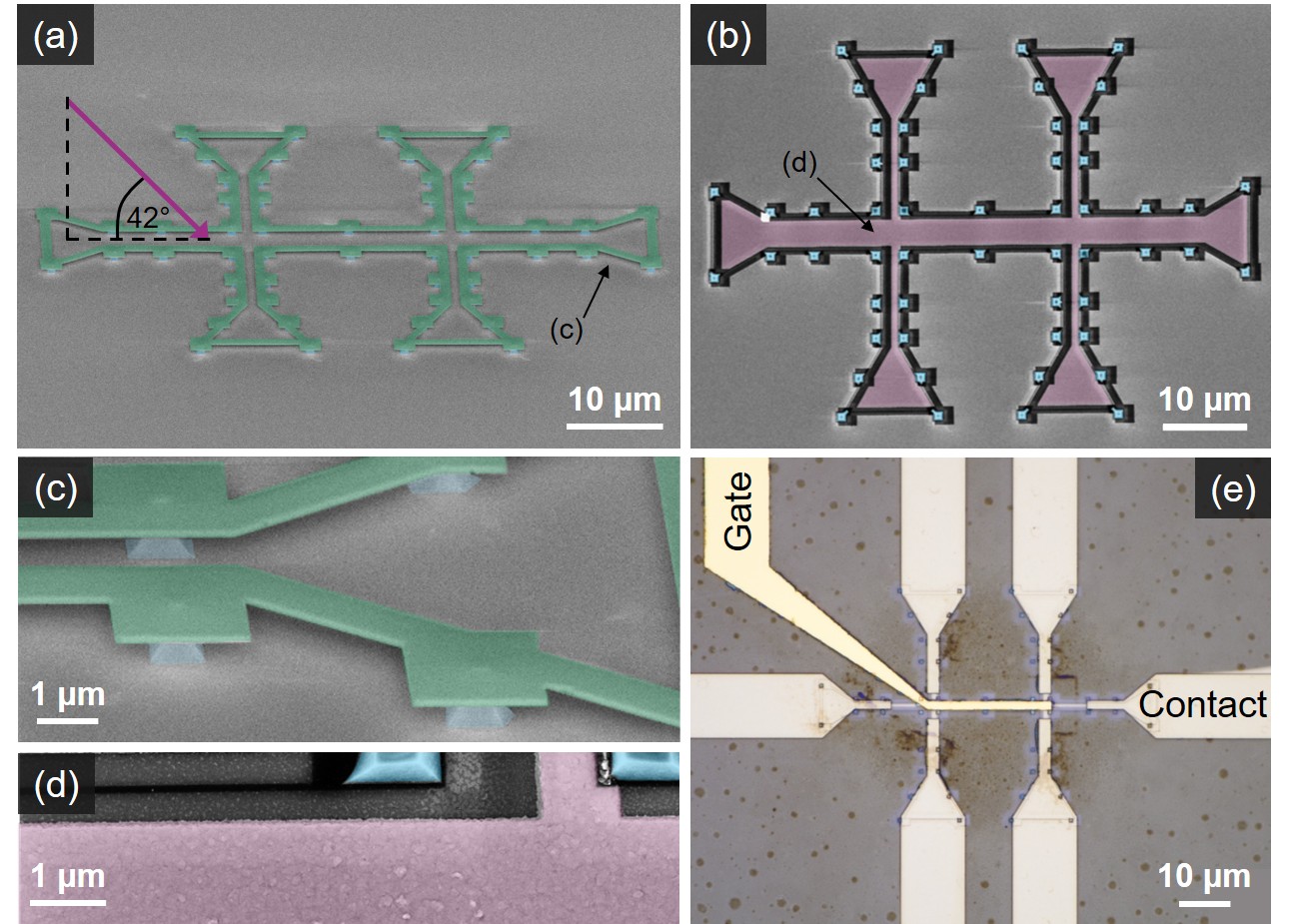}
\caption{Hall bar fabrication. (a) False colored SEM image of the stencil mask under an angle on the Si(111) growth substrate. The Si$_3$N$_4$ layer which defines the shape of the devices is depicted in green and the supporting SiO$_2$ pillars in light blue. The purple arrow indicates how the substrate was aligned for the digital alloying step so that the MBE deposition of all materials happens from a fixed direction in alignment with the mask and under a polar angle of 42$^\circ$. (b) False colored SEM image of a Hall bar device after MBE growth. After growth the Si$_3$N$_4$ layer is removed with a first contact polymer \cite{FirstContact2026} only the SiO$_2$ pillars remain. The isolated CBST film in Hall bar shape is depicted in purple. (c) Zoom into the image from (a). The mask is made out of multiple free standing Si$_3$N$_4$ bridges in between the SiO$_2$ pillars to enable the mask removal via the first contact polymer \cite{FirstContact2026}. (d) Zoom into the image from (b). The surface morphology of the CBST film becomes visible. (e) Optical microscopy image of the finished device after global etching of excess CBST, contact deposition (grey) and top gate deposition (gold).}
\label{Figure2}
\end{figure*}

Compared to the conventional MBE growth, where all materials are co-evaporated, for our employed digital alloying approach, all relevant materials are deposited sequentially. This technique has been previously used to precisely tune the composition of compound semiconductors~\cite{Miller1984,Cody1994,Kaspi2003} or to grow conventional topological insulator films from Bi$_2$Te$_3$ and Sb$_2$Te$_3$~\cite{Taylor2009,Peranio2012} and thin films of the intrinsic MTI MnBi$_2$Te$_4$~\cite{Gong2019}. In order to utilize this approach for our MTI micro- and nanostructures, stencil masks are fabricated onto the Si(111) substrate using SiO$_2$ and Si$_3$N$_4$ layers similar to the work by Sch\"uffelgen \textit{et al.}~\cite{Schueffelgen2019}. These stencil masks define the shapes of the MTI nanostructures, in our case Hall bar devices, during growth. The sequence of the growth process in the MBE chamber is illustrated in the diagram given in Fig.~\ref{Figure1}(a). After the substrate has been loaded into the MBE chamber it is degassed at $T_{\text{degas}}$ ($650\degree$C) to remove the hydrogen bonds formed during the previous hydrofluoric acid cleaning step from the Si(111) surface. The subsequent layer growth happens in cycles. Each cycle contains the following steps: First, the substrate is heated to T$_{\text{growth}}$ ($100$ - $110\degree$C). The cell of the first element, in this case Bi, is oriented at a specific azimuthal angle $\varphi$ relative to the substrate. Then, Bi is evaporated from this direction. Afterwards, the substrate is rotated so that the Cr cell has the same effective orientation to the substrate as the Bi cell. After depositing the Cr layer, the same process of rotating and deposition is repeated for Sb. Figures~\ref{Figure1}(b) and (c) show a schematic illustration of the Bi and Cr step, respectively. Then, the substrate temperature is increased to $T_{\text{anneal}}$ ($175\degree$C) during which Te is continuously evaporated from the same direction as the previous elements, forming the tetradymite crystal structure. Finally, Te deposition is stopped and the substrate is cooled back down to T$_{\text{growth}}$. In order to achieve a certain desired film thickness, this cycle can be repeated any number of times fully automatically. Increasing the thickness per cycle, a quantity that is introduced correspondingly to a conventional growth rate and measured in quintuple layers (QLs), from 1 QL to 1.25 QL leads to a better surface morphology but decreases the crystal quality. A low annealing temperature enhances crystal quality but also increases the formation of parasitic phases on the mask and misoriented grains. Careful balancing of these two parameters leads to the growth of high quality (Cr$_x$Bi$_y$Sb$_z$)$_2$Te$_3$ (CBST) thin films. In the following, $x$ will be referred to as the Cr content and represents the fraction of cation sites occupied by Cr, substituting Bi and Sb.

\subsection{Stencil mask fabrication and Hall bar contacting}

The stencil mask layout to fabricate the CBST Hall bars is shown in Fig.~\ref{Figure2}(a). The false-colored scanning electron microscopy (SEM) image shows the stencil on top of the Si(111) substrate. The Si$_3$N$_4$ layer which defines the shape of the devices is depicted in green and the supporting SiO$_2$ pillars in light blue. Figure~\ref{Figure2}(c) provides a closer look at the mask. It is evident that the stencil mask consists of a series of interconnected Si$_3$N$_4$ bridges, situated between the SiO$_2$ pillars. This allows to fully remove the Si$_3$N$_4$ layer using a first contact polymer \cite{FirstContact2026} and therefore facilitates contacting the Hall bars and enables the deposition of a top gate. Figure~\ref{Figure2}(b) shows a false-colored SEM image of a Hall bar device after digital alloying growth and subsequent mask removal. During growth, all cells were successively aligned with the direction indicated by the purple arrow in Figure~\ref{Figure2}(a). The CBST thin film in the shape of a Hall bar (highlighted in purple) is through this in-situ process fully isolated from the global layer of CBST outside of the mask (grey). A capping layer of 5\,nm of Al$_2$O$_3$ is deposited in-situ from all four cardinal directions so that the Hall bar is passivated. Figure~\ref{Figure2}(d) shows a close-up of the Hall bar device and reveals the surface morphology of the CBST film. In order to measure the Hall bar devices, the excess global layer of CBST is etched away while protecting the Hall bars with a photoresist and afterwards metal contacts and a top gate are deposited via lift-off processes. Figure~\ref{Figure2}(e) shows an optical microscopy image of a final device. Two main sets of Hall bar devices from (Cr$_x$Bi$_y$Sb$_z$)$_2$Te$_3$ are produced: Set A consists out of Hall bars with a width of $W=3\,\upmu$m and varying Cr contents of $x=0.019$, 0.032, 0.051, and 0.06. Set B Hall bars have a Cr content of 0.051 and widths including $W=3\,\upmu$m, $1\,\upmu$m, $500$\,nm, $300$\,nm and $100$\,nm. The Bi to Sb ratio is always roughly 1:1. The Cr contents were determined via Rutherford backscattering spectroscopy (RBS). The exact composition for each Cr content can be found in the RBS reports~\cite{Karthein-data2026}.

\subsection{Structural analysis}

\begin{figure}
\centering
\includegraphics[width=0.42\textwidth]{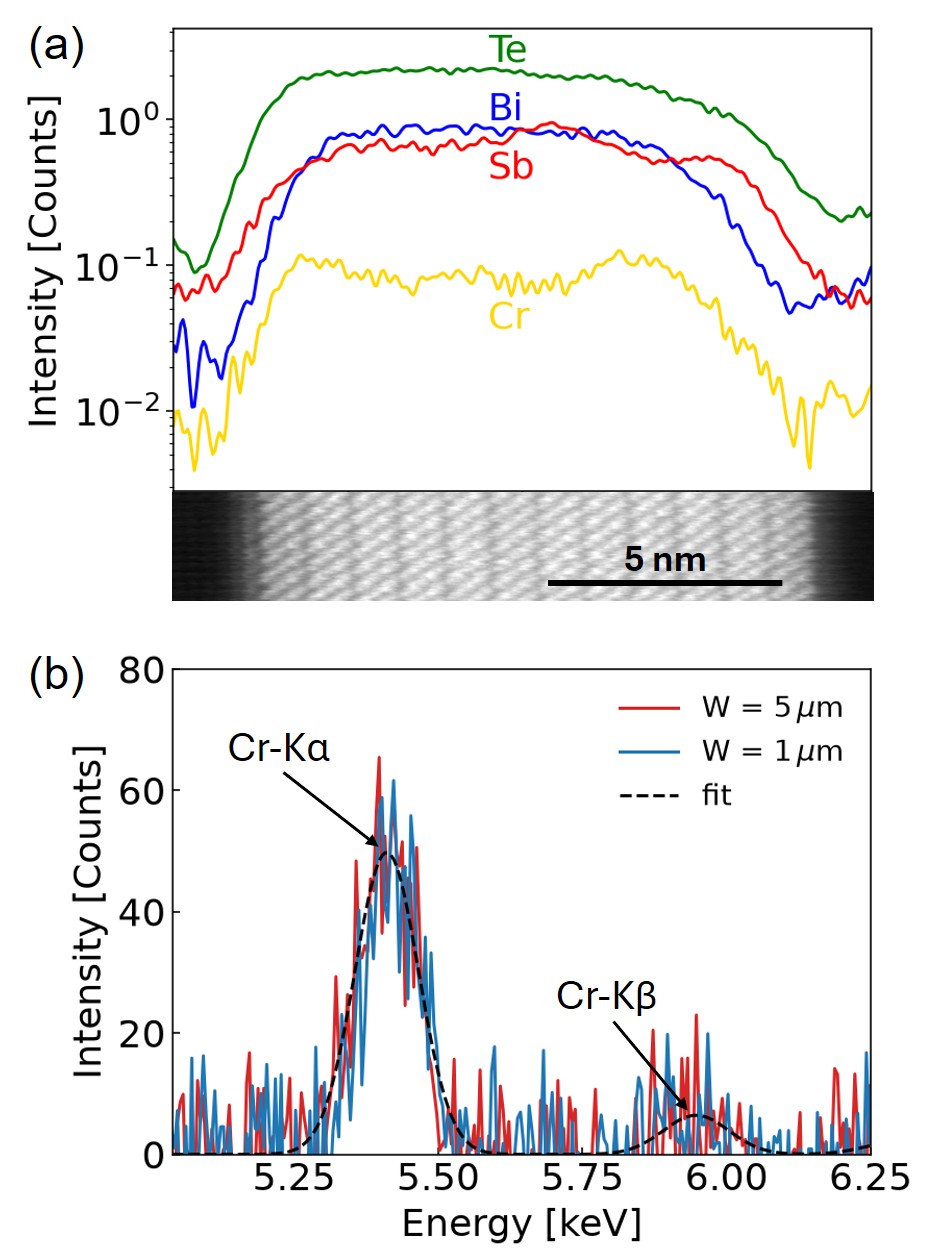}
\caption{Structural analysis of a previously grown sample. (a) EDX profile of a lamella cut from a Hall bar with a width of $W=1\,\upmu$m. The TEM image on the $x$-axis shows the quintuple layer structure clearly. (b) Comparison between EDX spectra of Hall bars with widths $W=5\,\upmu$m and $1\,\upmu$m in the energy region where the Cr peaks are expected. No qualitative difference between the spectra is found.}
\label{Figure3}
\end{figure}

To investigate the crystal structure, transmission electron microscopy (TEM) was used to analyse some comparable samples that were grown in the same way. Figure~\ref{Figure3}(a) shows the energy-dispersive x-ray (EDX) spectroscopy profile of a lamella cut from a 1-$\upmu$m-wide Hall bar made of (Cr$_{0.03}$Bi$_{0.47}$Sb$_{0.5}$)$_{2}$Te$_3$. The TEM image on the $x$-axis serves as a reference where the EDX profile was taken and reveals the quintuple layer structure of the MTI film. To check whether a narrower mask changes how many magnetic dopants are incorporated into the thin films, the lamella of the 1-$\upmu$m-wide device is compared with another lamella of a 5-$\upmu$m-wide one at the same composition. Figure~\ref{Figure3}(b) shows the EDX spectra of both lamella in the energy region where the Cr peaks are expected. It becomes clear that the peak height is very similar and therefore it is not expected that a more narrow masks prevents Cr atoms from reaching the devices. The most noticeable difference between the two lamellas is that the layer of the 5-$\upmu$m-wide Hall bar has a thickness of 14\,nm while the 1-$\upmu$m-wide structure has a smaller film thickness of 12.8\,nm. The thicknesses are determined from cross-sectional TEM images.

\begin{figure*}
\centering
\includegraphics[width=\textwidth]{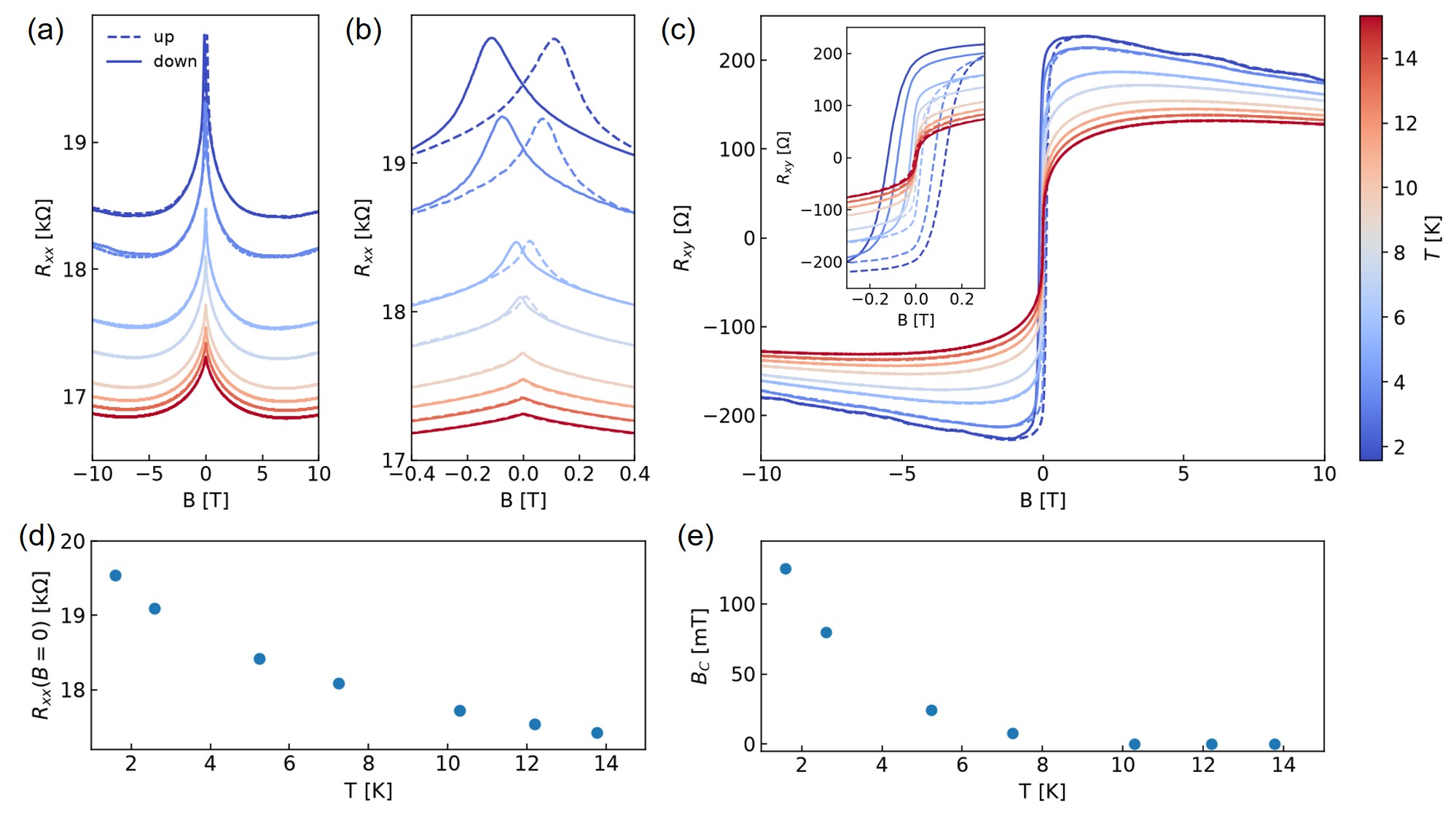}
\caption{Temperature dependent magnetotransport of a 3-$\upmu$m-wide Hall bar with a Cr content of 0.051. (a) $R_{xx}$ over the complete magnetic field range for temperatures ranging from $T=1.5$\,K to $14$\,K. (b) $R_{xx}$ for different temperatures in the low magnetic field range. A hysteresis is visible between the up-sweep (dashed line) and down-sweep (solid line) of the magnetic field. (c) $R_{xy}$ for varying temperatures in a magnetic field range of $\pm 10$\,T. At low fields (inset) the hysteresis is clearly visible. d) Temperature dependence of $R_{xx}$ at $B=0$. The decreasing trend implicates semiconducting behaviour. e) Coercive field $B_c$ against temperature. The coercive field is extracted from $R_{xy}(B_c) = 0$. The decrease of $B_c$ for increasing $T$ is expected for ferromagnetic materials.}
\label{Figure4}
\end{figure*}

\section{Magnetotransport Measurements}

To investigate the low-temperature magnetotransport properties, all devices from sets A and B are measured in a variable temperature insert (VTI) cryostat with a base temperature of 1.5\,K. First, the temperature-dependent (gate-dependent) magnetotransport measurements and analysis are presented for an exemplary sample with a specific Cr content of 0.051 (0.032) and a width of $W=3\,\upmu$m ($W=1\,\upmu$m). To study the effect of different Cr contents, the magnetotransport on Hall bars with the same dimensions but varying Cr contents are compared (set A). Further on, for a fixed Cr content the transport properties of Hall bars of different width are studied (set B).  

\subsection{Temperature dependent effects}

\begin{figure*}[t]
\centering
\includegraphics[width=\textwidth]{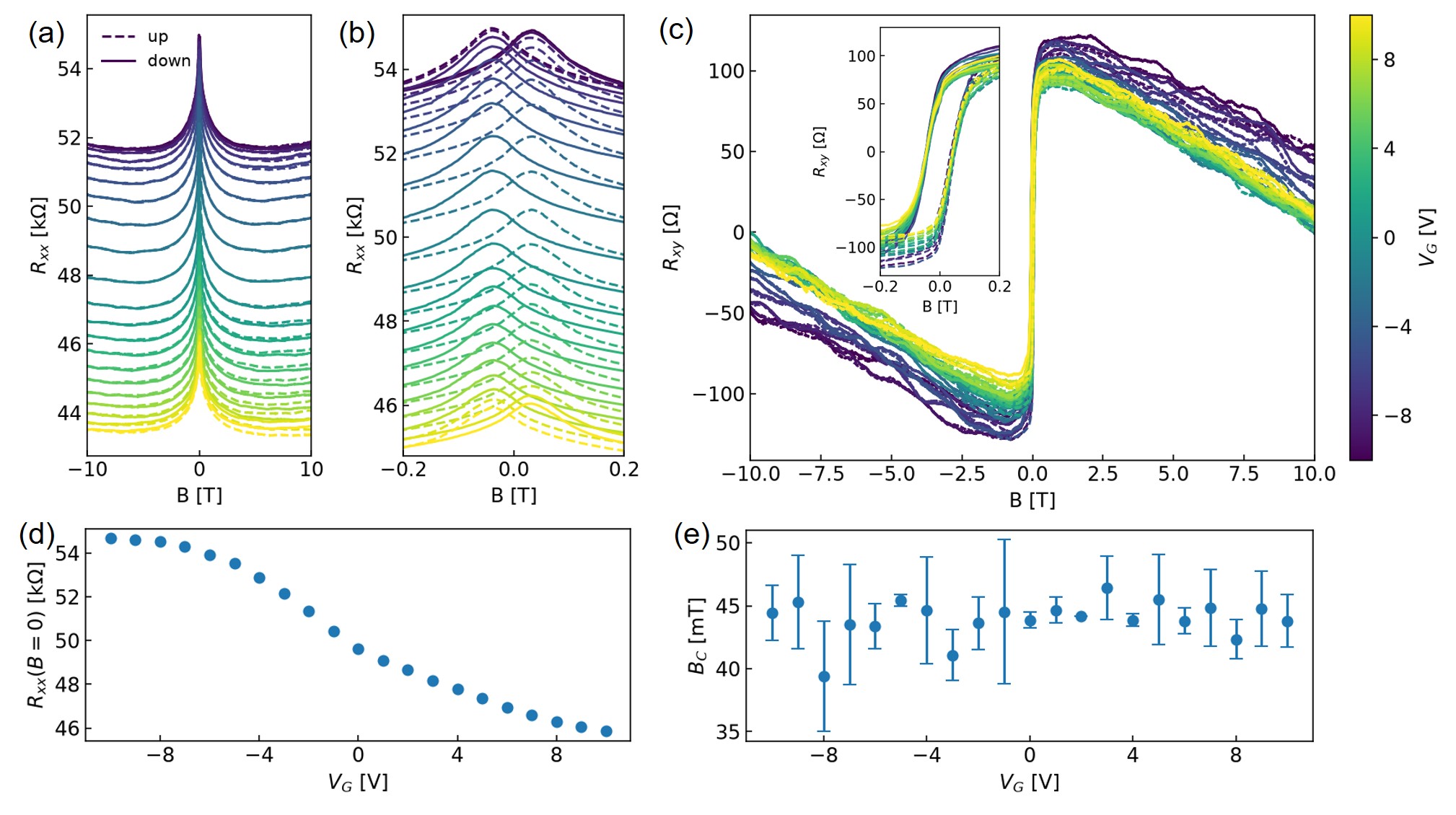}
\caption{Gate voltage dependent magnetotransport measurements of a 1-$\upmu$m-wide Hall bar with a Cr content of 0.032. (a) and (b) Longitudinal magnetoresistance $R_{xx}$ for gate voltages between $V_G=-10\,$V and $10\,$V over a large range of $B$-field and in a small range around zero field, respectively. (c) $R_{xy}$ for different gate voltages. The inset shows a zoom-in of the hysteresis. (d) The zero field resistance $R_{xx}(B=0)$ behaves monotonically as a function of gate voltage. (e) The coercive field $B_c$ as a function of $V_G$ stays basically constant.}
\label{Figure5}
\end{figure*}

First, we investigate temperature related effects by measuring the longitudinal $R_{xx}$ and Hall resistance $R_{xy}$ for all Hall bars in a wide range of temperatures. Figure~\ref{Figure4} shows such a set of measurements at the example of a Hall bar with a width of $W=3\,\upmu$m and a Cr content of 0.051. In Fig.~\ref{Figure4}(a) the longitudinal resistance curves over the complete magnetic field range of $\pm10$\,T is shown for temperatures from 1.5 to 14\,K. One finds that $R_{xx}$ decreases with increasing temperature, indicating insulator behaviour. The trend of the zero field value of $R_{xx}$ against temperature is shown in Fig.~\ref{Figure4}(d). Such a temperature dependence has been previously reported for strongly disordered CBST films, where the Fermi energy is located within the bulk band gap~\cite{Guo2021}. Figure~\ref{Figure4}(b) reveals the hysteretic nature of the longitudinal resistance for opposite directions of magnetic field sweeps. $R_{xx}$ exhibits a peak at the negative and positive coercive fields $\pm B_c$, which decreases with temperature. This becomes more apparent from the Hall resistance data shown in Fig.~\ref{Figure4}(c). The inset clearly shows how the hysteretic part of $R_{xy}$ slowly disappears at higher temperatures. The coercive field determined from $R_{xy}(\pm B_c)=0$ is plotted as a function of temperature in Fig.~\ref{Figure4}(e). Between 8 and 10\,K $B_c$ vanishes, which marks the transition from the ferromagnetic to the paramagnetic state, assuming that $R_{xy}$ is proportional to the magnetization $M$ of the sample. In reality, coercivity is governed by magnetization-reversal mechanisms, including domain-wall pinning, magnetic anisotropy, and thermal activation, rather than solely by the existence of ferromagnetic order~\cite{Strnat1988,Vertesy2002,Rong2006}. Therefore, $B_c$ vanishing can not be directly associated with the transition from ferromagnetic to paramagnetic behaviour in the magnetization $M$. Still, the loss of coercivity in $R_{xy}$ can be qualitatively discussed since we are interested in the magneto-transport properties of our MTI thin films. Above 10\,K the sample is behaving paramagnetic in $R_{xy}$, which is apparent from the non-linearity at low magnetic fields in Fig.~\ref{Figure4}(c). At high temperatures the shape of $R_{xy}$ at high magnetic fields is largely affected by approaching the magnetization of the layer. Therefore, it is not possible to extract the projected two-dimensional charge carrier density $n_\mathrm{2d}$ at elevated temperatures from the slope of $R_{xy}$~\cite{Ohno1998,Berry2001}. Later on, $n_\mathrm{2d}$ will be discussed only from measurements taken at sufficiently low temperatures at 1.5\,K, where the magnetization is saturated already. Additionally, the magnitude of the anomalous Hall effect is far from being quantized at $h/e^2 \approx 25.8\,\text{k}\Omega$. This is a result of bulk and surface transport dominating over the chiral edge channel. This together with the temperature dependence of $R_{xx}$ hints towards strong local fluctuations in the size of the magnetic gap as well as the presence of charge puddles which both cause the contribution of the chiral edge channel to be small compared to surface and bulk transport.~\cite{Chong2020,Guo2021,Lippertz2022}.

\subsection{Gate voltage dependent effects}

\begin{figure*}
\centering
\includegraphics[width=0.90\textwidth]{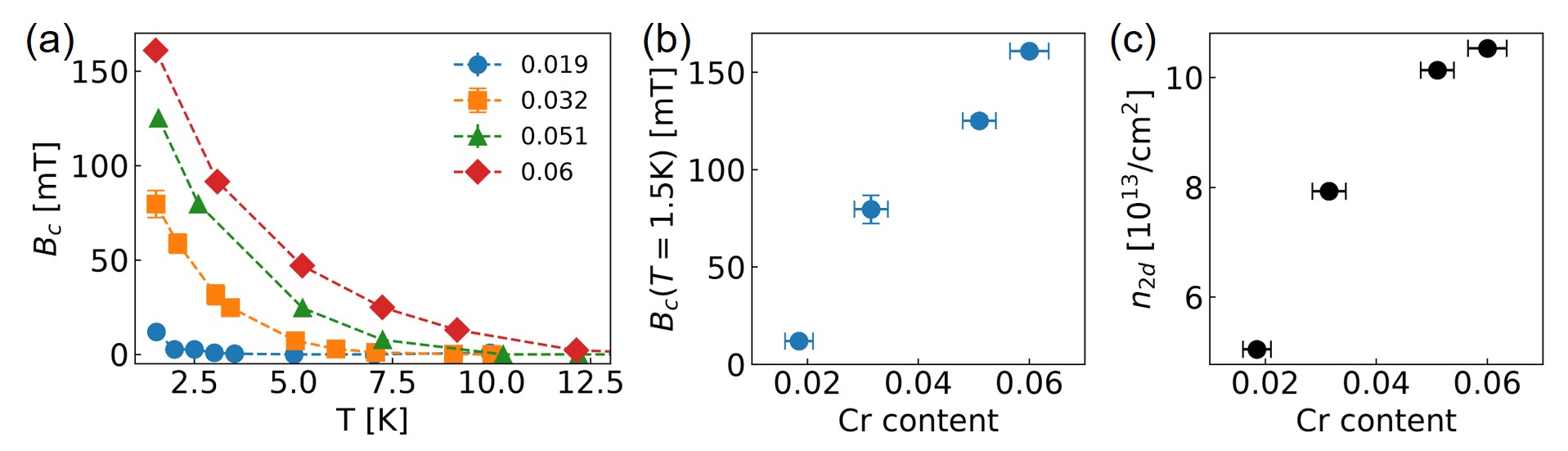}
\caption{Comparison of relevant transport parameters of 3-$\upmu$m-wide Hall bars for various Cr contents $x$. (a) $B_c$ versus $T$ for Cr contents from $x=0.019$ to 0.06. (b) $B_c$ at 1.5\,K against Cr content. (c) Electron concentration at $n_\mathrm{2d}$ at 1.5\,K as a function of $x$.}
\label{Figure6}
\end{figure*}

Next, the effect of a gate voltage on the magnetotransport properties of a Hall bar from set B with a width of $1\,\mu$m and a Cr content of 0.032 is studied. Figures~\ref{Figure5}(a) and (b) show $R_{xx}$ as a function of magnetic field for different gate voltages $V_G$. The measurements were taken at the base temperature of 1.5\,K. The longitudinal resistance at zero field changes monotonically as a function of $V_G$, which is also depicted more explicitly in Fig.~\ref{Figure5}(d). Since positive gate voltages cause a decrease of $R_{xx}(B=0)$ whereas negative voltages lead to a higher zero field resistance, the main charge carrier type contributing to transport can be assumed to be electrons. This assumption is consistent with the slope of the Hall voltage in Fig.~\ref{Figure5}~(c) at large magnetic fields being negative. Another noticeable feature is the modulations of $R_{xy}$ as a function of $B$ for higher magnetic fields, as well as a relatively small change of the ordinary Hall slope for varying gate voltages. Together with the generally low value of $R_{xy}$ they again hint towards high charge disorder and the presence of electron and hole charge puddles~\cite{Qiu2023,Lippertz2022}, in contrast to a compensation of $p$- and $n$-doping which would lead to a large value of $R_{xy}$.  A similar behaviour was already observed for conventionally grown CBST films on silicon substrates in our previous work~\cite{Karthein2025}. For non-magnetic materials, a two-channel model can be employed to extract the charge carrier densities for electrons and holes separately~\cite{Ito2022}, but in our case the presence of the ferromagnetic hysteresis makes this impossible~\cite{He2021}. Therefore, the charge carrier densities extracted from a regular Hall analysis might not produce the true value. The coercive field is mostly constant as a function of $V_G$, as shown in Fig.~\ref{Figure5}(e). The error bars stem from determining the positive and negative coercive field and averaging them. Gate voltage independent ferromagnetic properties, such as $B_c$, hint towards the charge carrier independent Van Vleck mechanism being mainly responsible for ferromagnetic order in our films~\cite{Chang2013,Kou2013}. 

\subsection{Effect of chromium content}

Next, we investigate the effect of different Cr contents. Therefore, set A consisting of 3-$\upmu$m-wide Hall bars with $x=0.019$, 0.032, 0.051, and 0.6 are compared. The raw data of all set A samples can be found in the Supplemental Material \cite{Karthein-suppl2026}. Figure~\ref{Figure6}(a) shows the $B_c$ vs. $T$ dependence at each Cr content, respectively. It becomes clear that the coercive field becomes zero at lower temperatures for lower values of $x$. Figure.~\ref{Figure6}(b) shows $B_c$ at base temperature plotted against the Cr content and reveals a linear increase. The results given in Figs.~\ref{Figure6}(a) and (b) are expected from a classical ferromagnetism standpoint and consistent with previous studies on the same material~\cite{Kou2013,Chang2013,Ou2016,Ye2015}. In Figure~\ref{Figure6}(c) the electron concentration $n_\mathrm{2d}$ at 1.5\,K is plotted as a function of Cr content. We found in our previous study~\cite{Karthein2025} that Cr acts as a $p$-type dopant in Cr-doped (Bi$_y$Sb$_{1-y}$)$_2$Te$_3$. The same result was also found by others~\cite{Gultom2024}. In contrast, in Fig.~\ref{Figure6}(c) we find the opposite behaviour where $n_\mathrm{2d}$ increases with increasing Cr content. A possible explanation is the formation of charge puddles where similar to our study on conventionally MBE-grown CBST thin films the classical Hall analysis breaks down and gives misleading results due to the compensation of electrons and holes~\cite{Karthein2025}. 

\subsection{Effect of Hall bar dimensions}

\begin{figure*}
\centering
\includegraphics[width=0.9\textwidth]{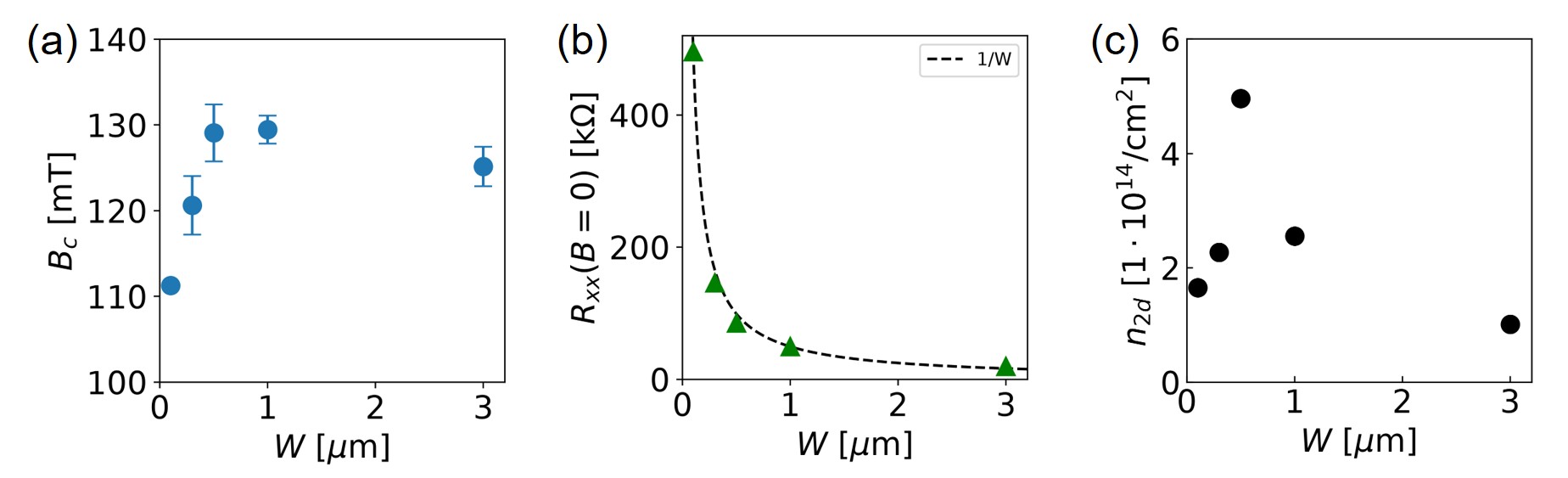}
\caption{Comparison of Hall bars with $x=0.051$ Cr content and varying widths $W$. (a) $B_c$ as a function of Hall bar width $W$. (b) Zero field resistance $R_{xx}(B=0)$ as a function of $W$. The black dashed line represents the expected $1/W$ dependence of $R_{xx}(B=0)$ (c) Electron concentration $n_\mathrm{2d}$ as a function of $W$.}
\label{Figure7}
\end{figure*}

Finally, set B Hall bars with a fixed Cr content of 0.051 and widths varying from $W=3\,\upmu$m down to $100\,$nm are compared. Figure~\ref{Figure7}(a) shows a basically constant $B_c$ with a small trend of lower values towards smaller Hall bar widths. This trend is possibly due to the thickness reduction of narrower Hall bars found from the TEM analysis. As can be seen in Fig.~\ref{Figure7}(b), the zero-field resistance of the Hall bars $R_{xx}(B=0)$ decreases with increasing width $W$. This decrease follows a $1/W$ dependence (dashed black line), indicating a basically constant resistivity, independent on the sample width. This is also in line with the electron concentration $n_\mathrm{2d}$ shown in Fig.~\ref{Figure7}(c), which was extracted from the slope of the Hall signal at large magnetic fields, i.e. where the magnetization due to the Cr-dopants is saturated \cite{Ohno1998}. One finds that $n_\mathrm{2d}$ remains in range of about $1\times10^{14}$ to $5\times10^{14}$\,cm$^{-2}$. The value itself seems to fluctuate a bit from device to device, which can be expected for strongly disordered thin films.

\section{Conclusion}

We successfully fabricated micro- and nanometer-sized Hall bars from Cr-doped (Bi$_y$Sb$_{1-y}$)$_2$Te$_3$ using digital alloying and stencil-mask techniques. A structural analysis of two Hall bar devices via TEM and EDX confirmed this through the crystal structure and the incorporated materials. The analysis also eliminates the suspicion that the stencil-mask hinders the deposition of materials for narrow structures. Subsequently, Hall bars with different Cr contents and a range of different widths are investigated in magneto-transport measurements. First, the temperature dependence of a specific Hall bar is discussed and implications of disorder and charge puddles are found from the temperature dependence of $R_{xx}$. Furthermore, the transition from ferromagnetic to paramagnetic behaviour in $R_{xy}$ with increasing temperature is observed and discussed with respect to the difference between coercivity in a Hall measurement and general ferromagnetic order. At the example of another Hall bar, gate voltage dependent measurements are presented and further implications for a disordered potential landscape are found. This is supported by contradictory trends of $n_\mathrm{2d}$ for increasing Cr contents. The coercive field and the transition temperature from ferromagnetic to paramagnetic transport signatures behave as expected if more Cr atoms are incorporated into the crystal. For a fixed Cr content, no significant width dependence of the coercive field is found and the sheet resistance $R_{xx}$ follows the expected $1/W$ trend. The electron concentration remains in the same order of magnitude and is generally not expected to be dependent on the Hall bar dimensions. Some fluctuations in the value can be again attributed to disorder.

\section*{Acknowledgments}

We thank Herbert Kertz for technical assistance. This work is supported by the QuantERA grant MAGMA and by the Deutsche Forschungsgemeinschaft (DFG, German Research Foundation) under Grant No. 491798118. In addition, we received funding from DFG under Germany’s Excellence Strategy - Cluster of Excellence Matter and Light for Quantum Computing (ML4Q) EXC 2004/2 – 390534769 as well as financial support by the Bavarian Ministry of Economic Affairs, Regional Development and Energy within Bavaria’s High-Tech Agenda Project "Bausteine für das Quantencomputing auf Basis topologischer Materialien mit experimentellen und theoretischen Ans\"atzen" (grant no. 07 02/686 58/1/21 1/22 2/23). J.K., M.K., and T.S. would like to express their gratitude for the support provided as part of the ASPIRE project JPMJAP2338.

\section*{Data Availability}

The data that support the findings of this article are openly available \cite{Karthein-data2026}.

\putbib[bu1.bbl]
\end{bibunit}

\cleardoublepage
\widetext

\titleformat{\section}[hang]{\bfseries}{\MakeUppercase{Supplemental Note} \thesection:\ }{0pt}{\MakeUppercase}
\begin{bibunit}[]
\setcounter{section}{0}
\setcounter{equation}{0}
\setcounter{figure}{0}
\setcounter{table}{0}
\setcounter{page}{1}

\renewcommand{\thesection}{\arabic{section}}
\renewcommand{\theequation}{S\arabic{equation}}
\renewcommand{\thefigure}{S\arabic{figure}}
\renewcommand{\figurename}{Supplemental Figure}
\renewcommand{\tablename}{Supplemental Table}
\renewcommand{\bibnumfmt}[1]{[S#1]}
\renewcommand{\citenumfont}[1]{S#1}

\begin{center}
\textbf{\large Supplemental Material: Digitally-alloyed magnetic topological insulator nanostructures grown via stencil mask molecular beam epitaxy}
\end{center}

\section{Surface morphology}

\begin{figure*}
\centering
\includegraphics[width=0.8\textwidth]{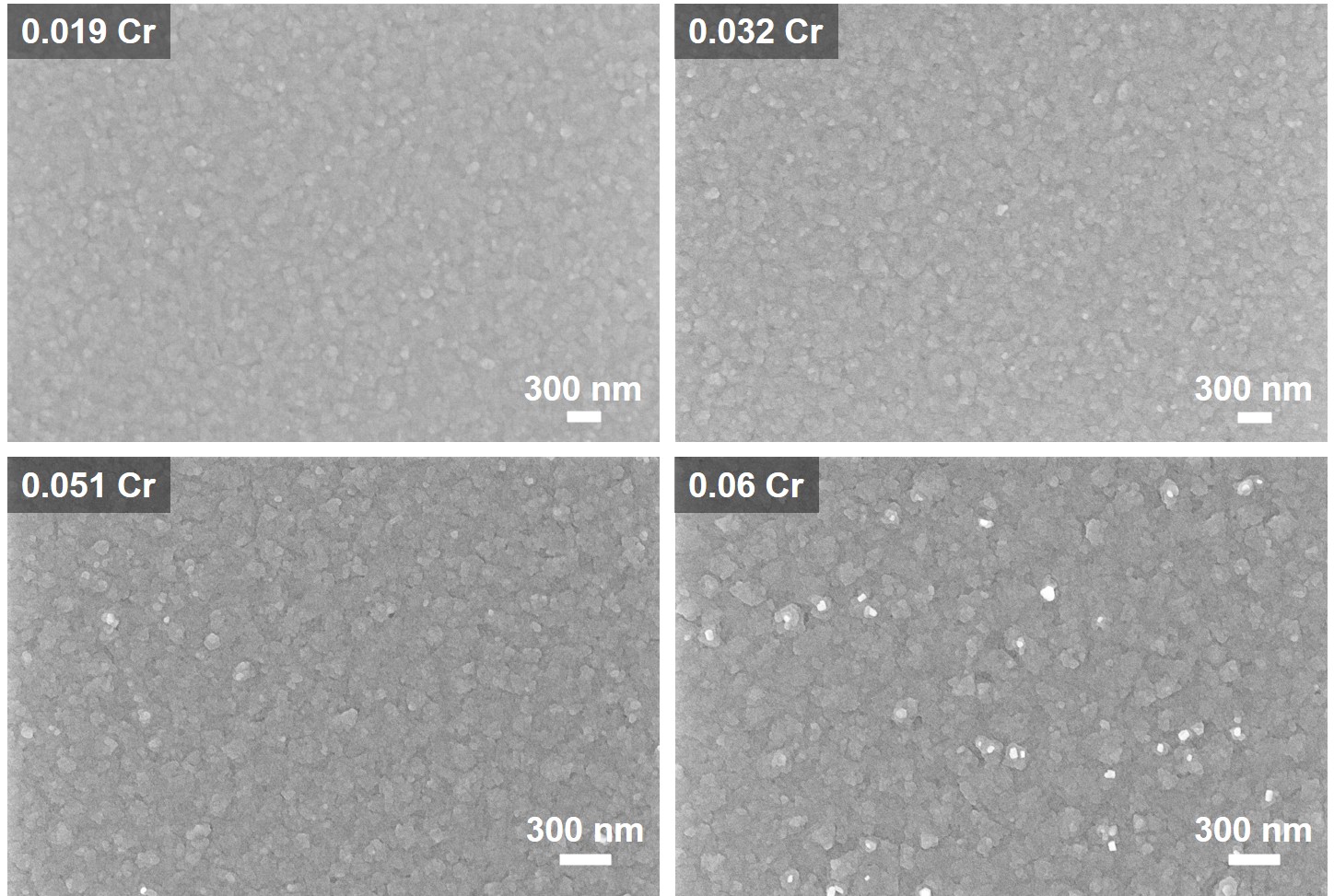}
\caption{SEM images of a CBST thin films at each Cr content that is discussed in the main text.}
\label{FigureS1}
\end{figure*}

Supplemental Figure~\ref{FigureS1} shows scanning electron microscopy (SEM) images of four Cr-doped (Bi Sb)$_2$Te$_3$ (CBST) thin films grown via digital alloying with varying Cr contents of 0.019, 0.032, 0.051, and 0.06. From these SEM images one finds that the surface becomes slightly rougher for increasing Cr content, which is in line with our previous study on CBST thin films with different Cr contents~\cite{Karthein2025}.

\section{Raw data of samples}

\subsection{Set A}

\begin{figure*}
\centering
\includegraphics[width=0.75\textwidth]{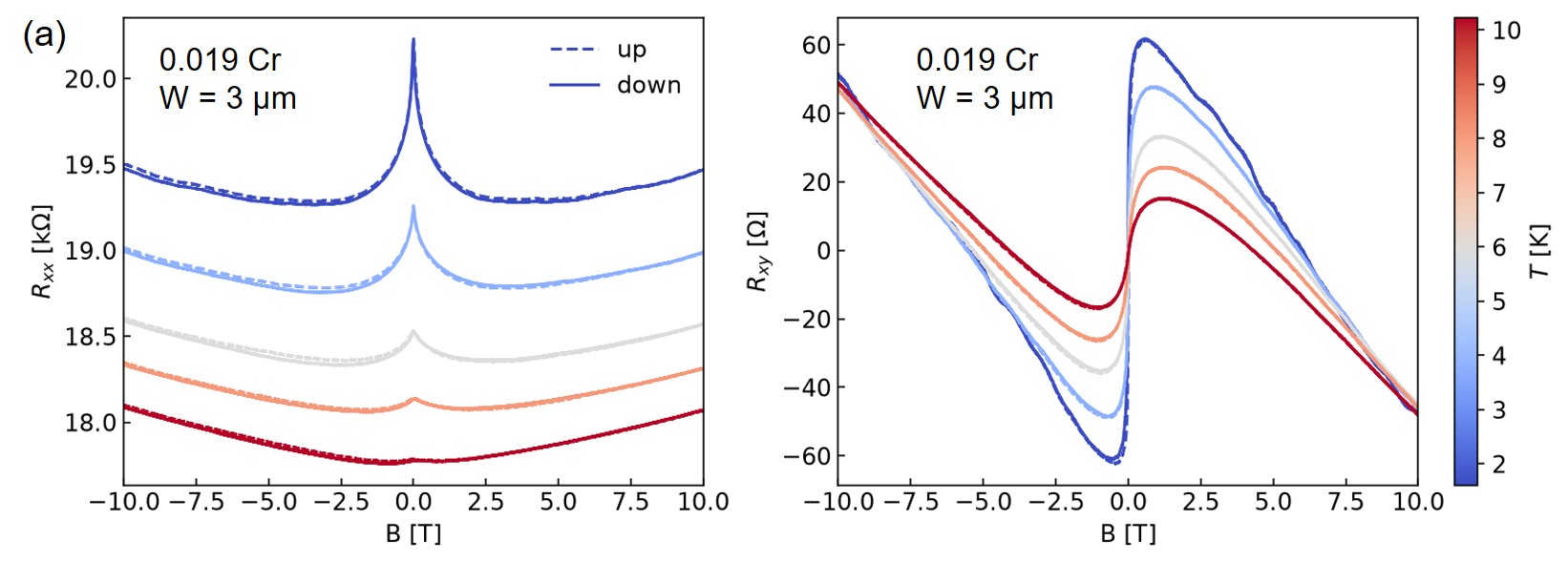}

\includegraphics[width=0.75\textwidth]{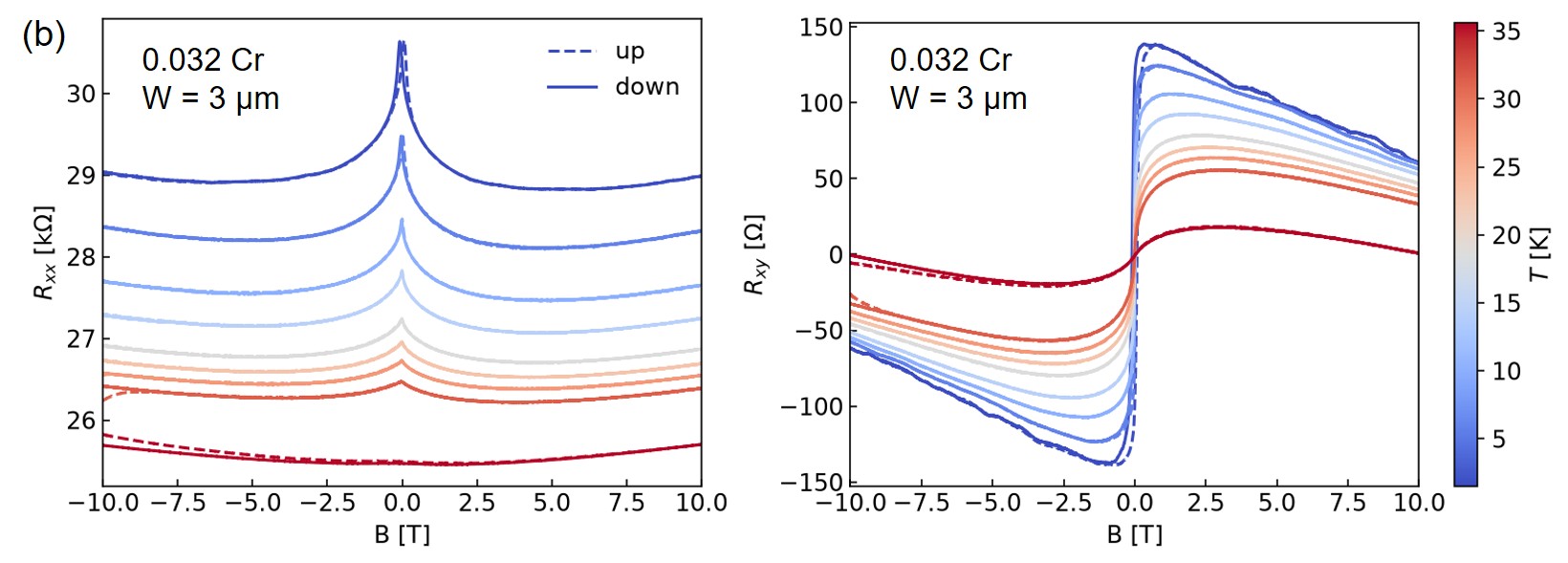}

\includegraphics[width=0.75\textwidth]{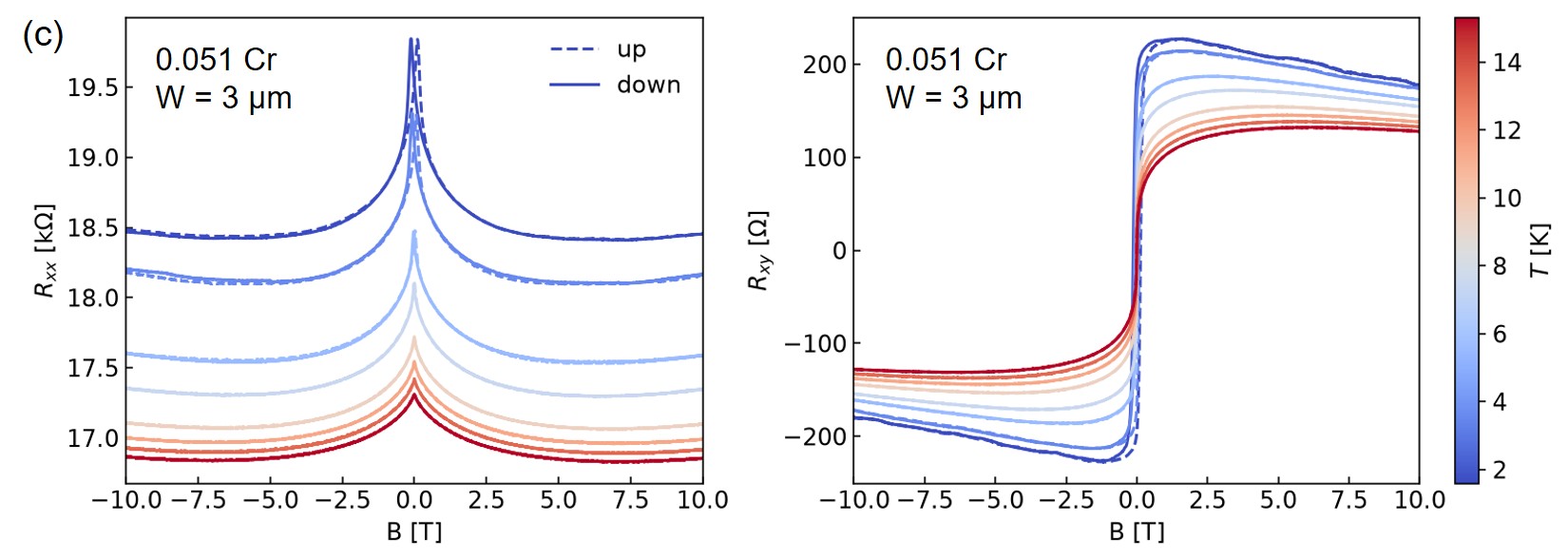}

\includegraphics[width=0.75\textwidth]{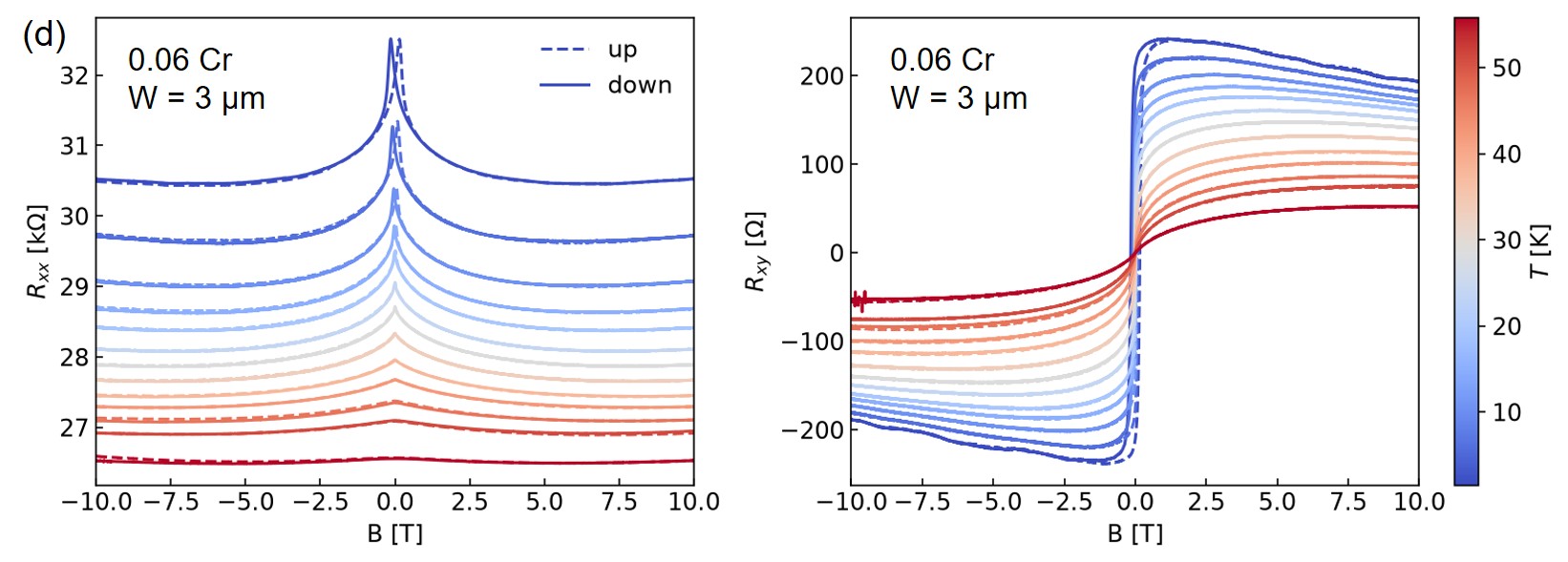}
\caption{Raw data of set A. Hall bars with a width of $3\,\upmu$m are measured for different Cr contents ranging from 0.019 to 0.06.}
\label{FigureGroupA}
\end{figure*}

Supplemental Figure~\ref{FigureGroupA} presents the raw data of the samples that are referred to as set A in the main text. Set A contains Hall bars with a width of $3\,\upmu$m and Cr contents of 0.019, 0.032, 0.051, and 0.06. All devices are measured for temperatures from 1.5\,K up to at least 10\,K or higher over a large magnetic field range between -10\,T to 10\,T. For each device $R_{xx}$ shows a decreasing trend with increasing temperature like the device shown in the main text. From the curve of $R_{xy}$ we extract the coercive field $B_c$ as a function of temperature for each Cr content as well as the charge carrier density at 1.5\,K.

\subsection{Set B}

\begin{figure*}
\centering
\includegraphics[width=0.75\textwidth]{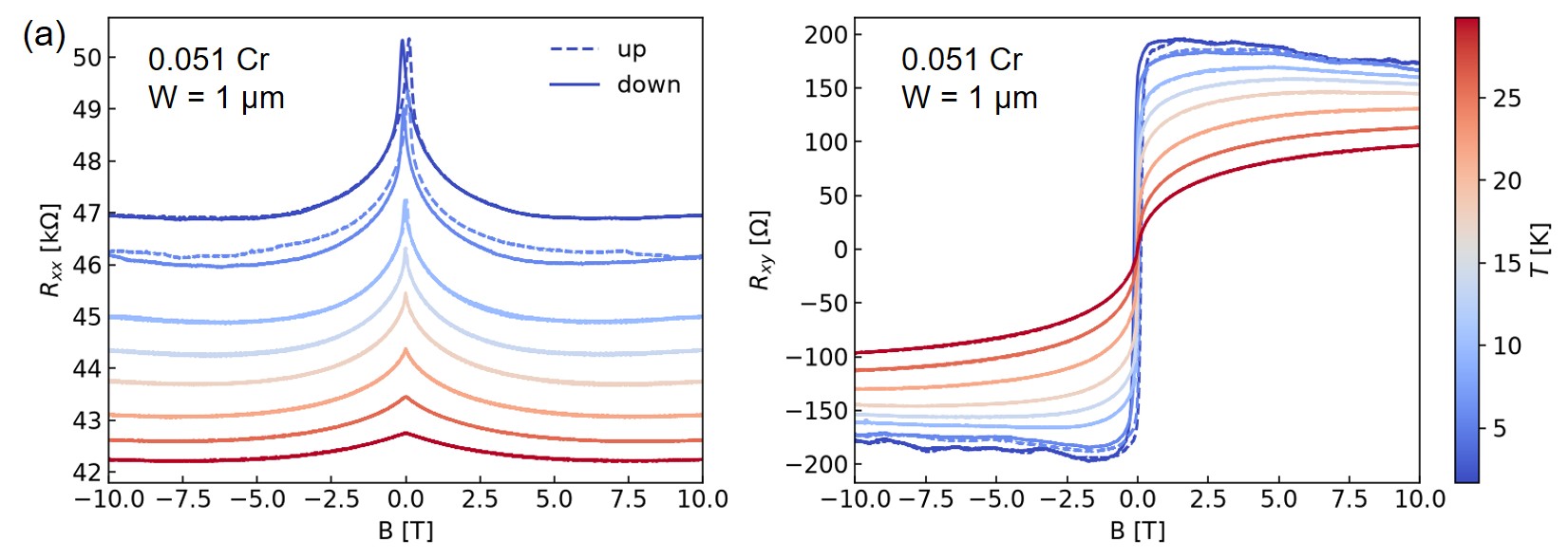}

\includegraphics[width=0.75\textwidth]{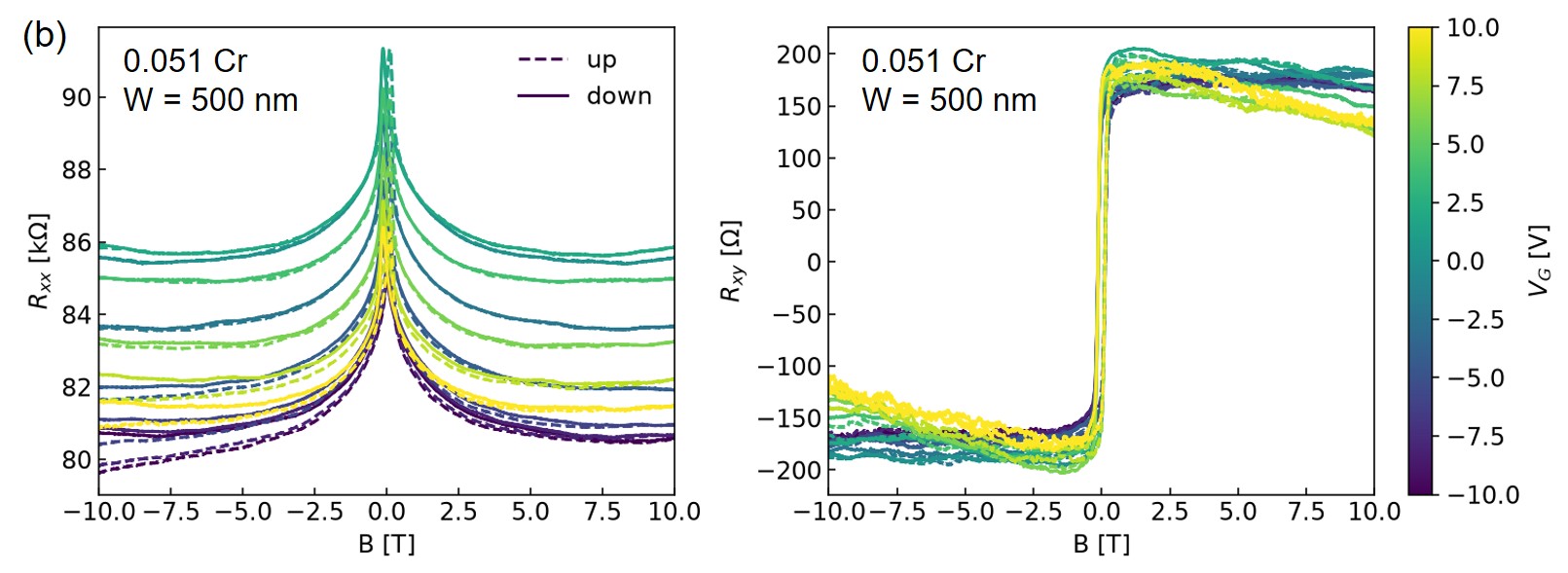}

\includegraphics[width=0.75\textwidth]{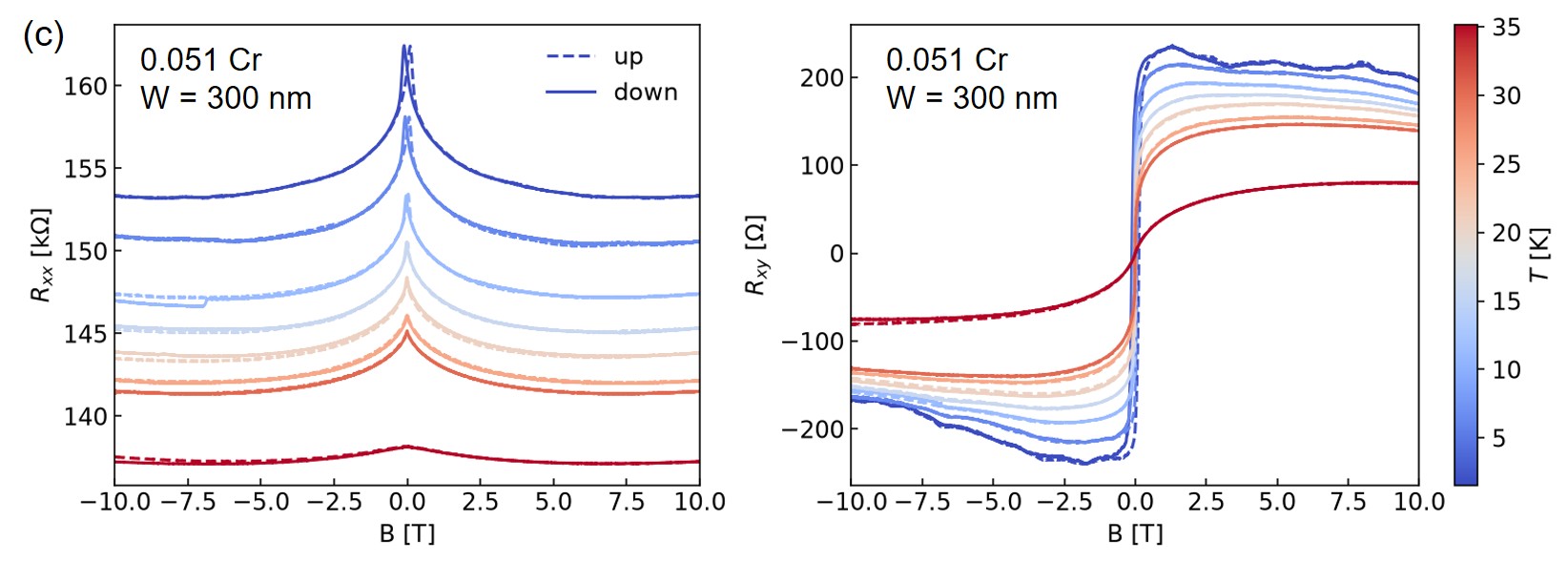}

\includegraphics[width=0.75\textwidth]{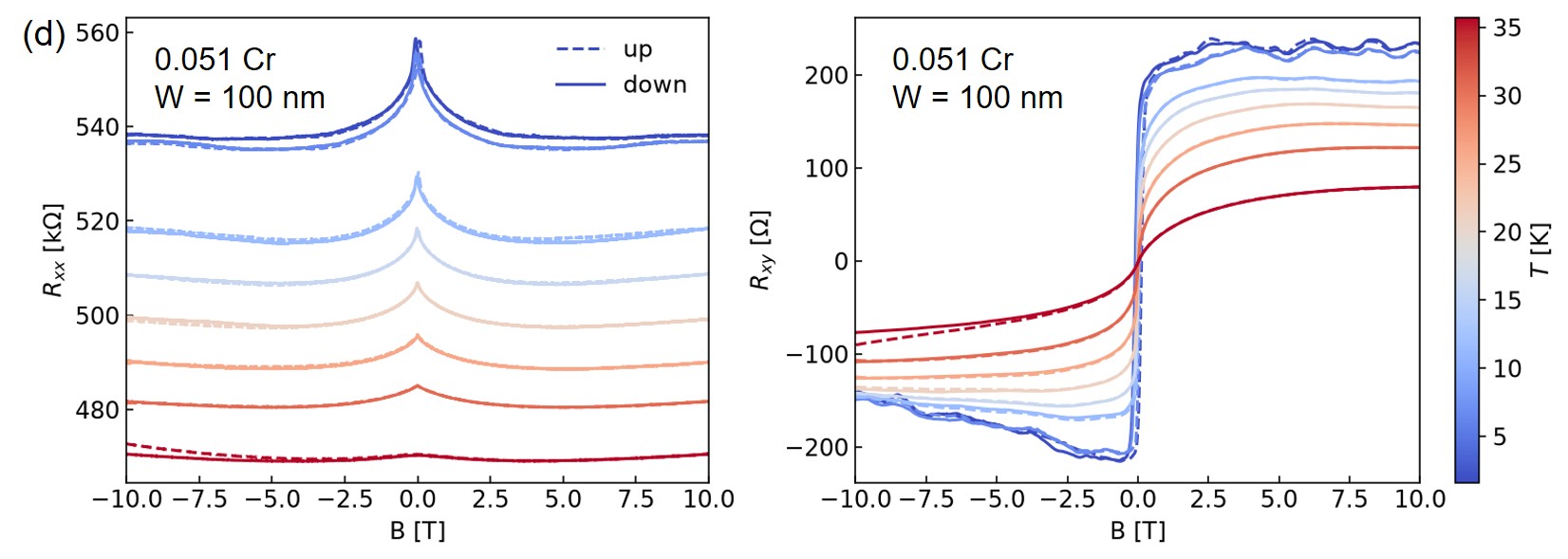}
\caption{Raw data of set B. Hall bars with a Cr content of 0.051 are measured for different Hall bar widths $W$.}
\label{FigureGroupB}
\end{figure*}

Set B contains Hall bars with a Cr content of 0.051 and widths including $3\,\upmu$m, $1\,\upmu$m, $500$\,nm, $300$\,nm, and $100$\,nm. Supplemental Figure~\ref{FigureGroupB} once again shows the raw data of $R_{xx}$ and $R_{xy}$ for all Hall bars in this set. The Hall bar with a width of $3\,\upmu$m and 0.051 Cr content is not shown again as it is already included in the data for group A. Supplemental Figure~\ref{FigureGroupB}(b) shows the curve for different gate voltages since high field data for this device was not recorded for different temperatures. From set B samples we extract the charge carrier density at 1.5\,K (at zero gate voltage) at high magnetic fields above 3\,T. In the sample shown in Supplemental Figure~\ref{FigureGroupB}(b), $R_{xx}$ at a constant field also does not show a clear gate dependence as the sample from the main text. This could be due to charge puddles~\cite{Qiu2023,Lippertz2022,Guo2021}, i.e. electron and hole contributions, being affected differently by a gate voltage, causing $R_{xy}$ to change non-monotonically.

\putbib[bu2.bbl]
\end{bibunit}

\end{document}